\documentclass[12pt]{article}

\DeclareRobustCommand{\rchi}{{\mathpalette\irchi\relax}}
\newcommand{\irchi}[2]{\raisebox{\depth}{$#1\chi$}} 

\begin{document}

\title{On the Role of a Torsion-like Field in a Scenario for the Spin Hall Effect.}
\maketitle
\begin{center}
\author{C. F. L. Godinho$^{a}$  \footnote{crgodinho@gmail.com}},
\author{J. A. Helay\"{e}l Neto$^{b}$  \footnote{josehelayel@gmail.com}}\\
{$^a$Departamento de F\'{\i}sica, Universidade Federal Rural do Rio de Janeiro (UFRRJ),
BR 465-07, 23890-971, Serop\'edica, RJ, Brazil\\
$^b$Centro Brasileiro de Pesquisas F\' isicas (CBPF),
Rua Dr. Xavier Sigaud 150, Urca, 
22290-180, Rio de Janeiro, Brazil \\}
\end{center}
\begin{abstract}
\noindent 
Starting from a field theory action that describes a Dirac fermion, we propose and analyze a model based on a low-relativistic Pauli equation coupled to a torsion-like term to study Spin Hall Effect (SHE). We point out a very particular connection between the modified Pauli equation and the (SHE), where what we refer to torsion as field playing an important role in the spin-orbit (SO) coupling process.  In this scenario, we present a proposal of a spin-type current, considering the tiny contributions of torsion in connection with geometrical properties of the material.  
\end{abstract}

\pagestyle{myheadings}
\markright{Godinho, and Helay\"{e}l Neto}

\pagestyle{myheadings}
\section{Introduction}
Today, the technological advances provide a wide spectrum of manipulations with the spin degrees of freedom. These amazing and reliable kinds of procedure are propelling the spintronics as a consolidated sub-area of Condensed Matter Physics \cite{Wolf}.  Since the experimental advances are increasingly provi\-ding high-precision data, many theoretical works are being presented \cite{Culcer,Mu,Yao,Mish} for a deeper understanding of the phenomenon.  The understanding of the spin currents inside semiconductors is an important field of research with remarkable applications.

 Its importance, considering the scientific and commercial aggregate value,  is bringing more interest and new ideas have been presented ever since.  We also point out some gauge-field theoretic approaches to the Physics of Graphene \cite{Dayi}, Lagrangians formulations with an $SU(2)$ gauge symmetry associated to spin forces \cite{Tao} and the quantized spin Hall conductivity \cite{Yao2}.

The connection between spin-orbit interaction and spin currents was theoretically predicted by Diakonov and Perel \cite{Dya} and, subsequently, by Hirsh \cite{Hi}.  Experimentally, its confirmation has recently been found \cite{Ka,Wu}.  The eventual occurrence  of impurities in the presence of spin-orbit coupling is currently known as Extrinsic Spin Hall Effect, and ever since this subject has raised up a great deal of attention among physicists. 

In a more recent stage, two independent groups \cite{Mu,Si} have proposed the Intrinsic Spin Hall Effect, based on peculiarities of the semiconductor structure band.  Despite these well-consolidated definitions, it appears that there is some cloud after all.    Even though there are differences between both kinds of phenomena, some authors seem to strongly believe that they are the same phenomenon, but with different mechanisms for the charge coupling \cite{Dya2,Wang}.  

Indeed, we realize that the task here is to try to explain a very complex and singular low relativistic level quantum-mechanical phenomenon, for which many theoretical approaches have been presented.  For High Energy Physics, if any fundamental theory cannot be verified, one might try some approach via effective theories that could take us closer to some particular understanding of lower-energy effects.  Adopting Relativistic Quantum Mechanics basics as guide line, our work sets out to present an alternative point of view for understanding the mobility of two-dimensional electron systems (2DES), approaching the Spin Hall Effect as a geometrical consequence of spin-orbit coupling also generated by a field that we identify as a torsion-type degree of freedom.  

The torsion field  appears as one of the most natural extensions of General Relativity; along with the metric tensor, which couples to the energy-momentum distribution, torsion  inspects the details of the spin density tensor.  Actually, in General Relativity, fermions naturally couple to torsion by means of their spin. Electromagnetism is associated to charges, currents and magnetic dipoles; gravity is produced by mass. A torsion field is generated by the spin density of a given microscopic or macroscopic system.  Even though torsion is usually presented in connection with General Relativity, our description introduces torsion in connection with the geometry of the crystal \cite{Kat}, as it shall be put in more details later on.    \'Elie Cartan was the first to point out the important role of torsion in solids \cite{Cart}. Riemann-Cartan geometry, where the torsion appears as a key geometrical object, is a topic of particular interest for condensed matter physics, especially in connection with the theory of defects in solids \cite{Kat}.  We point out the paper \cite{Kou} where the authors develop a very interesting study of the role of torsion on conduction electron scattering in solids, see also \cite{Aur}.   

Our main purpose here is to present a low-relativistic description for the SHE by adopting an action for Dirac spinors, and terms that yield CPT- and -Lorentz symmetry breaking.   Besides, we shall realize that the Lagrangian density has a covariant derivative given by $D_{\mu}=\partial_{\mu}+i\eta \gamma_5 S_{\mu}+ieA_{\mu}$, with a torsion pseudotrace $(S_{\mu})$  contribution term, responsible for the dyna\-mics of the electrons in the geometry of the crystal, whereas $A_{\mu}$ describes the lattice electromagnetic potential. Here, the dynami\-cs of the electron follows the non-relativistic limit of Dirac equation.  The torsion is directly connected with the fixed geometry of the solid; for this reason, we treat it a non-dynamical entity. The new term $\eta$ is a coupling parameter introduced in \cite{Sha1}, which for our purposes, measures the interaction of the electrons to some sort of defect represented by the background field or some geometrical property from the sample energy bands.    Considering that effects of torsion have never been measured, it is a fascinating idea to be able to make concrete predictions about its effects, and how to measure them \cite{Sha1a,Hamm1}.

After  presenting the introduction our work is outlined as follows: in Section 2 we  did a brief discussion about the torsion, we present some relevant references about it and consider its role in our approach.  In section 3 we obtain an extended Pauli equation, starting from Dirac's usual action with a torsion sector.  In Section 4 we consider a phenomenological analysis where SHE is taken in account with the Pauli equation.   Finally in Section 5 we display our final considerations and future perspectives.
\section{General Aspects on Torsion}
Why is it so necessary to talk about torsion?  Most of times, it is a concept misinterpreted or forgotten.  In a general-space time,  we find two different entities: curvature and torsion.  Following Einstein's assumptions,  energy and momentum or merely matter, produces the space-time curvature, while Cartan was the first to investigate the importance of the torsion in the gravitation.  In fact, the Einstein-Cartan theory presents both curvature and torsion where, in the same way matter is the source of curvature and spin is the source of torsion space-time.  According to this theory, curvature and torsion are representing independent degrees of freedom of the gravitational field, and we can in this scenario presuppose a physics associated with torsion, in a macroscopic level where spins vanish, it coincides with general relativity, but in microscopic whenever spins are relevant we must consider torsion effects, some interesting sources for a good review about this theme are \cite{Sha2,Hehl,Hamm}.

In physics, it is very usual to make use of effective models whenever a fundamental theory is still unknown or is not well understood.    What happens when unusual properties of space-time are involved? How can we handle such new phenomena?  Indeed we may find the answer in of the classical gravitation, by merely considering the concept of torsion.  We intend here to show some of the main aspects of the torsion field.     The development of Gravitation is realized with the concept of covariant derivative, $\nabla_{\beta} X^{\alpha}=\partial_{\beta}X^{\alpha}+\Gamma_{\beta\gamma}^{\alpha}A^{\gamma}$; in this case, for a contravariant vector. Usually the affine connection is considered as a symmetric quantity : $\Gamma_{\beta\gamma}^{\alpha}=\Gamma_{\gamma\beta}^{\alpha}$, and the metricity postulate or condition $\nabla_{\alpha} g_{\mu\nu}$ is a common place most of times.  These choices impose only a unique solution for $\Gamma_{\beta\gamma}^{\alpha}$, given by the simplest affine connection, the Christoffel symbol:
\begin{equation}
\Gamma_{\beta\gamma}^{\alpha}=\left\{{}^\alpha_{\beta\gamma} \right\}={1\over2} g^{\alpha \kappa}(\partial_{\beta}g_{\kappa\gamma}+\partial_{\gamma}g_{\kappa\beta}-\partial_{\kappa}g_{\beta\gamma})\,.
\end{equation}
However in a more general setting, where the geometry is not fully described by the metric tensor, in this case it becomes necessary to introduce an extra term to obtain an adequate physical solution,
\begin{equation}
\tilde{\Gamma}_{\beta\gamma}^{\alpha}=\left\{{}^\alpha_{\beta\gamma} \right\}+K^{\alpha}_{\beta\gamma}\,.
\end{equation}
The new piece , $K^{\alpha}_{\beta\gamma}$ , is skew-symmetric in $\beta\gamma$, is referred to as contorsion tensor \cite{Hehl}, and the torsion is nothing but:
\begin{eqnarray}
T^{\alpha}_{\beta \gamma}={1\over2}\left(\Gamma{^{\alpha}_{\beta \gamma}-\Gamma{^{\alpha}_{\gamma \beta}}}\right)\,.
\end{eqnarray}
In this scenario, we find motivation to pursue an investigation on the effects of torsion in connection with the SHE and SO-coupling.  Thought a crystal does not display continuous translational and rotational symmetries, if we are interested to study excitation limit the long wavelenght (low frequency) it becomes sensible to adopt a field-theoretic description on the continuous and Riemann-Cartan geometry appears as very systematic approach \cite{Kleinert}.
In our proposal, we shall focus on the torsion degrees of freedom and their coupling to fermions described by Dirac's equation.  Actually, according to Lorentz covariance torsion $T_{\alpha \beta \gamma}$, can be split  into three irreducible components, where one of them is called the pseudo-trace $S^{\kappa}={1\over6}\epsilon^{\alpha \beta \gamma \kappa}T_{\alpha \beta \gamma}$ .  How\-ever, as well-known, only its pseudo-vetor component couples to space-time fermions.  In our particular case, electron is the fermion we are considering.  So, as it shall become clear in the sequel, we are coupling torsion through its pseudo-vector component to the electron and possible contributions of this interaction to the Spin Hall Effect.  Actually, the scenario we propose in connection with torsion is that its pseudo-vector piece is not dynamical; it rather plays the role of a background tensor that measures some geometric anisotropy of the sample, this shall become more clear in the next sections.

\section{Pauli Equation with Torsion}
Dirac's equation is formulated to describe relativistic quantum mechanics properties of massive charged spin-1/2 particles.  Though the electrons in solid or more generally condensed matter systems are not relativistic $(v/c \approx 1/300)$, tiny relativistic effects should be taken into consideration as high-precision measurements are available through a number of nanotechnological apparatuses.  So one may adopt the view point of Dirac's equation to systematically derive, in the non-relativistic limit, correct terms to the dynamics of electrons in condensed matter systems that, otherwise would be introduced by hand.  Dirac's formulation naturally guides us to end up with relevant relativistic effects that appear as precision corrects in the low (Fermi) velocity regime.

Let us start off by considering the following action in natural $(c=\hbar=1)$ units,  
\begin{eqnarray}
\label{torsion action}
S=\int d^4x \bar{\psi}  \big( i\gamma^{\mu}D_{\mu} \,+\, \lambda \Sigma^{\mu\nu}\partial_{\mu}S_{\nu}\,+\,m \big){\psi};
\end{eqnarray}
this action describes the Dirac fermion non-minimally coupled to the background. Though the crystal has a lattice structure, we are focusing our attention on large-wave limit excitations; large in comparison with the lattice spacings, so that a continuous (field-theoretical) description is adopted and the torsion, that represents the geometry of the crystal and its anisotropy, is taken as a field itself.  This field has the status of a fixed background, for we assume the lattice geometry to be fixed. Our sort of gravity background does not exhibit metric fluctuations neither curved space-time, then the space-time is taken to be flat, and we propose a scenario such that the type of gravitational background is parameterized by the torsion pseudo-trace $S_{\mu}$, whose origin may be traced back to one geometrical defect or anisotropy in energy bands.  The energy bands in solids are directly connected with the potential energy for electrons in a periodic crystal and since electrons are subjected to quantum mechanical rules and they behave like waves, Pauli's exclusion principle guide us how these bands are filled by electrons. 

We would like to remark that we are considering only the $S_{\mu}$-torsion component because that is equivalent to consider the completely antisymmetric  component of torsion and this is the irreducible piece of torsion which couples to the fermions of the Standard Model, in our case, the electron.  From the action (\ref{torsion action}) the Dirac equation can be written as below:
\begin{eqnarray}
\big[i\gamma^{\mu}\partial_{\mu}-\eta\gamma^{\mu}\gamma_{5}S_{\mu}-e\gamma^{\mu}A_{\mu}+\lambda \Sigma^{\mu\nu} \partial_{\mu}S_{\nu}+m\big]\psi=0. \nonumber\, \\
\end{eqnarray}
We must emphasize here to avoid any misunderstanding that our purpose is not to re-invent the Dirac equation.  Our approach only aims to an extension of it, where the covariant derivative is now more sensible and capable to feel possible (geometrical) effects arising from the band structure of the samples under study. This includes the anisotropy of the crystal, that, as already mentioned, is described now by the torsion field.  Obviously, if such effects are weak or null we recover the usual equation.
\newpage
Defining the gauge-invariant momentum, $\pi_j=i\partial_j-eA_j$, the effective Hamiltonian takes the form
\begin{eqnarray}
\label{Hamiltonian}
H&=&\alpha_k\pi_k + \eta \gamma_5 S_0 -\eta \alpha_k \gamma_5 S_k + e A_0 - \nonumber \\
&+&{{i\lambda}\over{4}} \epsilon_{ijk}\beta \gamma_5 \alpha_k \partial_i S_j+\beta m\,.
\end{eqnarray}
 The semi-classical motion of electrons with applied electric and magnetic fields are given in the Heisenberg picture, with the position, $\vec{x}$, and momentum, $\vec{\pi}$, operators obey two different kinds of relations; we consider the torsion as a function of position only, $S=S(\vec{x})$, so that
\begin{eqnarray}
\dot{\vec{x}}&=&\vec{\alpha} \nonumber \\
\dot{\vec{\pi}}&=&e(\vec{\alpha} \times \vec{B})+e \vec{E}+\eta \gamma_5
{{\partial}\over{\partial x}}(\vec{\alpha} \cdot \vec{S})\vec{\hat{x}}
\,.
\end{eqnarray} 
One reproduces the usual relation for $\dot{\vec{x}}$, while the equation for $\dot{\vec{\pi}}$ presents a new term apparently giving some tiny correction to the Lorentz force.  

However, if we consider the torsion in a broader context, now as a momentum- and position-dependent background field, $S=S(\vec{x},\vec{k})$, we have to deal with the following picture,  

\begin{eqnarray}
\dot{\vec{x}}&=&\vec{\alpha}-\eta \gamma_5
{{\partial }\over{\partial k}}(\vec{\alpha} \cdot \vec{S})\vec{\hat{z}} \nonumber \\
\dot{\vec{\pi}}&=&e(\vec{\alpha} \times \vec{B})+e \vec{E}+\eta \gamma_5
{{\partial}\over{\partial x}}(\vec{\alpha} \cdot \vec{S})\hat{x}+ \nonumber \\
&+&e \eta \gamma_5  {{\partial A}\over{\partial x}}{{\partial}\over{\partial k}}(\vec{\alpha} \cdot \vec{S})\vec{\hat{z}}
\,. 
\end{eqnarray}
The two sets of dynamical equations above are clearly showing us the small corrections induced by the torsion term.  Nevertheless, we are still here in the  relativistic domain, and it is necessary change this framework for a better understanding of the SHE phenomenology.  For this reason, we are going to approach the system by going over into its non-relativistic regimen $(\mid \vec{p}\mid << m)$.
So, in this physical landscape, from now and hereafter, our goal is to consider a low-relativistic approximation based on an extended Pauli equation version by including torsion as presented before.  Employing the Hamiltonian (\ref{Hamiltonian}), we could carry out our calculations in the framework of the Fouldy-Wouthuysen transformations \cite{Sha4,Sha5}; however for the sake of our approximation at lowest order in $v/c$, we take that SHE is adequately well described by the low-relativistic Pauli equation.  We are considering that the electron velocities are in the range of Fermi's velocity.   

After working out  the usual non-relativist limit, we arrive at the version given below for the Pauli's equation:    
\begin{eqnarray}
\label{pauli0}
i{{\partial \varphi}\over{\partial t}}&=& \Big[{{(\vec{p}-e\vec{A})^2}\over{2m}}-{{e}\over{2m}}[(\vec{\sigma}+{{\eta}\over{2m}}\vec{S}+{{i\eta}\over{2m}}(\vec{\sigma}\times \vec{S}))\cdot \vec{B}]+ \nonumber \\
&+& eA_0 - \vec{\sigma} \cdot(\eta\vec{S}+ {{i\lambda}\over{4}}(\vec{\nabla}\times {\vec{S}}))- {{\lambda}\over{8m}}(\vec{\nabla}\times {\vec{S}})\cdot(\vec{\sigma}\times \vec{p})+ \nonumber \\ 
&+& {{i\lambda}\over{8m}}(\vec{\nabla}\times {\vec{S}})\cdot \vec{p}-{{e\lambda}\over{8m}}(\vec{\nabla}\times {\vec{S}})\cdot(\vec{\sigma}\times {\vec{A}})+ \nonumber \\
&+&{{ie\lambda}\over{8m}}(\vec{\nabla}\times {\vec{S}})\cdot \vec{A}\Big]\varphi\,. 
\end{eqnarray} 

The equation above displays the usual Pauli terms, but corrected by new terms due to the torsion coupling.  The second and the fourth contributions in the RHS of eq.(\ref{pauli0}) can be thought of as effective terms for $\vec{\sigma}$ and $\vec{S}$, respectively.  Then, by rewriting the equation, we now have:    
\begin{eqnarray}
\label{pauli1}
i{{\partial \varphi}\over{\partial t}}&=& \Big[{{(\vec{p}-e\vec{A})^2}\over{2m}}-{{e}\over{2m}}(\vec{\sigma}_{eff} \cdot \vec{B})+eA_0 - (\vec{\sigma} \cdot \vec{S}_{eff})+\nonumber \\
&-& {{\lambda}\over{8m}}(\vec{\nabla}\times {\vec{S}})\cdot(\vec{\sigma}\times \vec{p})+ {{i\lambda}\over{8m}}(\vec{\nabla}\times {\vec{S}})\cdot \vec{p}+\nonumber \\
&-&{{e\lambda}\over{8m}}(\vec{\nabla}\times {\vec{S}})\cdot(\vec{\sigma}\times {\vec{A}}) 
+{{ie\lambda}\over{8m}}(\vec{\nabla}\times {\vec{S}})\cdot \vec{A}\Big]\varphi\, 
\end{eqnarray}
where we considered a more simplified notation $\vec{S}_{eff}=\eta\vec{S}+ {{i\lambda}\over{4}}(\vec{\nabla}\times {\vec{S}})$ and $\vec{\sigma}_{eff}=\vec{\sigma}+{{\eta}\over{2m}}\vec{S}+{{i\eta}\over{2m}}(\vec{\sigma}\times \vec{S})$.
 
The fifth contribution in the RHS of eq. (\ref{pauli1}) is proportional to the Rashba SO-coupling term; this term affects the dynamics of the spin. This new term is naturally obtained in the non-relativistic regime of our Dirac equation in the presence of torsion, and it matches the relevance of the SO-coupling in the physics of the SHE.  We then emphasize that it is the torsion the responsible for the appearance of the SO-coupling in our model.
\subsection{A Particular Case}

Let us first propose a very particular case for a planar torsion field given by, $\vec{S}={1\over2}\rchi(x\hat{y}-y\hat{x})$; where $\rchi$ is a constant field-like quantity. This choice allows us to realize the curl of torsion as an intrinsic or effective magnetic field, $\vec{\nabla}\times {\vec{S}}=\vec{B}_{eff}=\rchi\hat{z}$.  According with our ansatz $\rchi$ behaves like a magnetic field, so $\vec{S}$ must be some kind of potential vector linked on a Berry phase like a flux through the sample, so that $$\gamma_{n}= \oint  d \vec{R}\cdot \vec{S}_n\,=\,\int {\rchi}_n \, dA \,. $$

In so doing, we may cast Pauli's equation in a simplified way,

\begin{eqnarray}
\label{pauli2}
i{{\partial \varphi}\over{\partial t}}&=& \Big[{{(\vec{p}-e\vec{A})^2}\over{2m}}-{{e}\over{2m}}(\vec{\sigma}_{eff} \cdot \vec{B}) - (\vec{\sigma} \cdot \vec{S}_{eff})+eA_0+ \nonumber \\
&-& {{\lambda \rchi}\over{8m}}\hat{z}\cdot(\vec{\sigma}\times \vec{p})-{{\lambda \rchi}\over{8m}}\hat{z}\cdot(e(\vec{\sigma}\times \vec{A})-i(\vec{p}+ e\vec{A}))\Big]\varphi\,. \nonumber \\
\end{eqnarray}

and, taking that ${\hat{x}}\cdot \vec{\sigma}={\hat{y}}\cdot \vec{\sigma}=0$ (the spins are aligned orthogonally to the plane of motion),

\begin{eqnarray}
\label{pauli4}
i{{\partial \varphi}\over{\partial t}}&=& \Big[{{(\vec{p}-e\vec{A})^2}\over{2m}}-{{e}\over{2m}}(\vec{\sigma}_{eff} \cdot \vec{B}) - (\vec{\sigma} \cdot \vec{S}_{eff})+eA_0 + \nonumber \\
&-& {{\lambda \rchi}\over{8m}}\hat{z}\cdot(\vec{\sigma}\times \vec{p})+{i{\lambda \rchi}\over{8m}}(\hat{z}\cdot \vec{p})\Big]\varphi\,.
\end{eqnarray} 
That is the final form for the Pauli's equation after the torsion coupling has been taken into account.  

\section{Spin Hall Effect, Spin Current \\ and Phenomenology}
It is very well understood that a direct consequence of low relativistic range, the spin and momentum states of an electron can be coupled.
The so called SO-coupling provides many experimental possibilities of studying electrons currents and some interesting approaches based on non-relativistic one particle, exhibiting SO-coupling can be found, describing SHE \cite{Si,Chu}. Considering once more our equation (\ref{pauli0}) but now in the limit $\vec{B_{ext}}= 0$ and $\vec{v} \perp \hat{z}$, where the conditions for Spin Hall Effect are being respected, we obtain the following Pauli equation
\begin{eqnarray}
\label{pauli0}
i{{\partial \varphi}\over{\partial t}}&=& \Big[{{\vec{p}\,^2}\over{2m}}+ eA_0 - \vec{\sigma} \cdot(\eta\vec{S}+ {{i\lambda}\over{4}}(\vec{\nabla}\times {\vec{S}}))- {{\lambda}\over{8m}}(\vec{\nabla}\times {\vec{S}})\cdot(\vec{\sigma}\times \vec{p})+ \nonumber\\ 
&+& {{i\lambda}\over{8m}}(\vec{\nabla}\times {\vec{S}})\cdot \vec{p}\Big]\varphi\, . 
\end{eqnarray} 
or, if we once more realize $(\vec{\nabla}\times {\vec{S}})=\rchi \hat{z}$ as an effective magnetic field, perpendicular to $\vec{v}$

\begin{eqnarray}
\label{pauli final}
i{{\partial \varphi}\over{\partial t}}&=& \Big[\underbrace{{{\vec{p}\,^2}\over{2m}}+eA_0}_{H_0}-\underbrace{(\vec{\sigma} \cdot \vec{S}_{eff})}_{H_{mag}} - \underbrace{{{\lambda }\over{8m}}\rchi \hat{z}\cdot(\vec{\sigma}\times \vec{p})}_{H_{SO}}\Big]\varphi\,.
\end{eqnarray}

The spin couplings are now regulated not only by the Pauli matrices, but they get corrections duel to the torsion term.
The torsion field can be realized as responsible for describing some sort of defect connected with the sample geometry and/or the impurities on it.  Consequently,  it is possible to propose a re-reading of the SHE within this approach. As shown above, our Pauli equation has basically the same structure as the usual in the kinetic and potential parts plus some corrections on the spin sector.  

The first and second terms in the RHS of eq. (\ref{pauli final}) correspond to those in the usual Schr\"odinger equation in the presence of some electric field ($H_0$).  The term responsible for describing interaction with the spin shows us a new kind of Zeeman sector given by  $(\vec{\sigma} \cdot \vec{S}_{eff})$; yielding corrections coming from the torsion and its curl ($H_{mag}$). Besides, our equation exhibits a SO-coupling term due to an inversion asymmetry of the confining potential; and there appears now a Rashba term $(\hat{z}\cdot(\vec{\sigma}\times \vec{p}))$, ($H_{SO}$) it shall be used to infer about the magnitude of the torsion, based on the non-relativistic limit approach.  

If we understand $({{\lambda \rchi}/{8m}})$ as the usual Rashba coupling constant, experimental data can be used to set up some bound on our constant $\lambda$. To estimate it, we recall that the magnitude for the Rashba coupling constant can be determined by experimental data, basically controlling the gate voltage \cite{Jun}.  Since we are working with natural units the Rashba coefficient is reached for InAs of $({{\lambda \rchi}/{8m}})=10^{-11}$ eV/m in \cite{Sch} and for our purposes we realize the torsion curl field $\rchi$ as an intrinsic or effective but not external magnetic field in a possible range of $\rchi \leq 20$ mT as a reasonable ansatz,  and then, our constant shall be constrained below as $\lambda= 9.322 \times 10^{-5}$MeV/T .

The Hamilton equations of motion can readily be derived from our Pauli Hamiltonian version and yields,
\begin{eqnarray}
\label{mov}
\dot{\vec{r}}&=&{\vec{k}\over m}- {{\lambda }\over{8m}}(\vec{\sigma}\times \vec{\rchi}), \nonumber \\
\dot{\vec{k}}&=&{1\over m}\Big(e\vec{E}+ \eta \vec{\nabla}(\vec{\sigma}\cdot \vec{S})\Big).
\end{eqnarray} 
The first equation is merely a kinematic equation of motion.  The second one, for charged particle with magnetic moment, but without a external magnetic field, is a dynamical equation, where the first term is the electric force, followed by a new term that can be understood as a force generated by the torsion contribution $\vec{f_t}=-\vec{\nabla}(\vec{\sigma}\cdot \vec{S})$.  This force gives an important contribution to the vector spin polarity. 

Let us now realize a practical experimental picture here, where we are considering the following set-up: of course $\vec{B}=0$, and $\vec{E}=E_z \vec{\hat{z}}$, $\vec{v}=(v_x,v_y,v_z)$, $\vec{\sigma}=(\sigma_x,\sigma_y,\sigma_z)$ and $\vec{S}={1\over2}\rchi(-y,x,z)$.  Within these limits, we obtain by means of equations (\ref{mov}), the following expressions for the $\dot{\vec{k}}$-components,
\begin{eqnarray}
\dot{k_x}&=&+{1\over2} \eta \rchi \sigma_x \nonumber \\
\dot{k_y}&=&-{1\over2} \eta \rchi \sigma_y \nonumber \\
\dot{k_z}&=&eE_z+{1\over2} \eta \rchi \sigma_z.
\end{eqnarray}
Even though in Spintronics there is not a general consensus on the concept of spin current \cite{Qin,Maekawa,Shi}, we understand, once $\vec{J}=\vec{L}+\vec{S}$, that it is very reasonable to realize the current as \cite{Mu}.  Following this idea, we can think of it as being induced by some electronic field in terms of the total angular momentum conservation, since that ${\dot{{J_z}}}=({\dot{\vec{r}}}\times \vec{k})_z+(\vec{r}\times{\dot{\vec{k}}})_z+{\dot{{k}}\vec{\hat{z}}}$.  In our case, the torque of the second term is zero too, because in our set-up $\dot{k_z}$ is parallel to $z$.  
\section{Concluding Remarks}
The main objective or our work is to propose a field theory model with torsion for studying the Spin Hall Effect and its eventual unfolds.
It is important to stress in our model that the SHE appears as a consequence of electrons moving in a crystal, but the mechanism underneath the effect is based on interaction terms that naturally emerge whenever we drive the (relativistic) Dirac equation to the non-relativistic limit.

We addressed our problem starting with such lagrangian field theory with a Dirac fermionic field plus a torsion term and subsequently we obtained the low relativistic limit of it.  We identified a natural torsion coupling with spin sector including the Rashba term and extra terms.   We believe we have obtained a good estimation for the spin current, $\dot{\vec{S}}$, despite torsion being a non-trivially detectable field yet.  The torsion field is giving us an interesting contribution that we understand, can be connected with some geometric aspects of the sample or perhaps with the impurities on it, as we had already anticipated in the introduction.  A possible interpretation of SHE is that the left and right chirality component of the electron are distinguished by the coupling to torsion.  The electro magnetic interaction alone, is not able to do that.

We are aware of the quantitative obstructions to suitably carry out experimental procedures for the  SHE.  Following this idea, we conclude by  hoping that some specific measurement procedure may be implemented to estimate a correlation between the geometrical aspects of the experimental sample and its consequences on the SHE observed.

\end{document}